\pdfoutput=1

\documentclass[11pt]{article}

\usepackage[final]{acl}

\usepackage{times}
\usepackage{latexsym}

\usepackage[T1]{fontenc}

\usepackage[utf8]{inputenc}

\usepackage{microtype}

\usepackage{inconsolata}

\usepackage{graphicx}

\usepackage{booktabs}
\usepackage{amsfonts}
\usepackage{amsmath,amssymb}
\usepackage{multirow}

\title{DIFFA: Large Language Diffusion Models Can Listen and Understand}

\author{
 \textbf{Jiaming Zhou\textsuperscript{1}$^*$}, \textbf{Hongjie Chen\textsuperscript{2}},  \textbf{Shiwan Zhao\textsuperscript{1}}, \textbf{Jian Kang\textsuperscript{2}}, \textbf{Jie Li\textsuperscript{2}}, \textbf{Enzhi Wang\textsuperscript{1}},
 \\
  \textbf{Yujie Guo\textsuperscript{1}}, \textbf{Haoqin Sun\textsuperscript{1}}, \textbf{Hui Wang\textsuperscript{1}}, \textbf{Aobo Kong\textsuperscript{1}}, \textbf{Yong Qin\textsuperscript{1}$^\dag$}, \textbf{Xuelong Li\textsuperscript{2}$^\dag$} 
\\
\\
     \textsuperscript{1} College of Computer Science, Nankai University, \\
     \textsuperscript{2} Institute of Artificial Intelligence (TeleAI), China Telecom, China,
\\
 \small{
   \textbf{Correspondence:} \href{mailto:zhoujiaming@mail.nankai.edu.cn}{zhoujiaming@mail.nankai.edu.cn}, \href{mailto:qinyong@nankai.edu.cn}{qinyong@nankai.edu.cn}
 }
}

\begin{document}
\maketitle
\renewcommand{\thefootnote}{}
\footnotetext{$^*$ This work was done during an internship at TeleAI.}
\footnotetext{$^\dag$ Yong Qin and Xuelong Li are corresponding authors.}
\begin{abstract}
Recent advances in large language models (LLMs) have shown remarkable capabilities across textual and multimodal domains. In parallel, diffusion-based language models have emerged as a promising alternative to the autoregressive paradigm, offering improved controllability, bidirectional context modeling, and robust generation. However, their application to the audio modality remains underexplored. In this work, we introduce \textbf{DIFFA}, the first diffusion-based large audio-language model designed to perform spoken language understanding. DIFFA integrates a frozen diffusion language model with a lightweight dual-adapter architecture that bridges speech understanding and natural language reasoning. We employ a two-stage training pipeline: first, aligning semantic representations via an ASR objective; then, learning instruction-following abilities through synthetic audio-caption pairs automatically generated by prompting LLMs. Despite being trained on only 960 hours of ASR and 127 hours of synthetic instruction data, DIFFA demonstrates competitive performance on major benchmarks, including MMSU, MMAU, and VoiceBench, outperforming several autoregressive open-source baselines. Our results reveal the potential of diffusion-based language models for efficient and scalable audio understanding, opening a new direction for speech-driven AI. Our code will be available at \url{https://github.com/NKU-HLT/DIFFA.git}.
\end{abstract}

\begin{figure}[!t]
\centering
\includegraphics[width=0.40\textwidth]{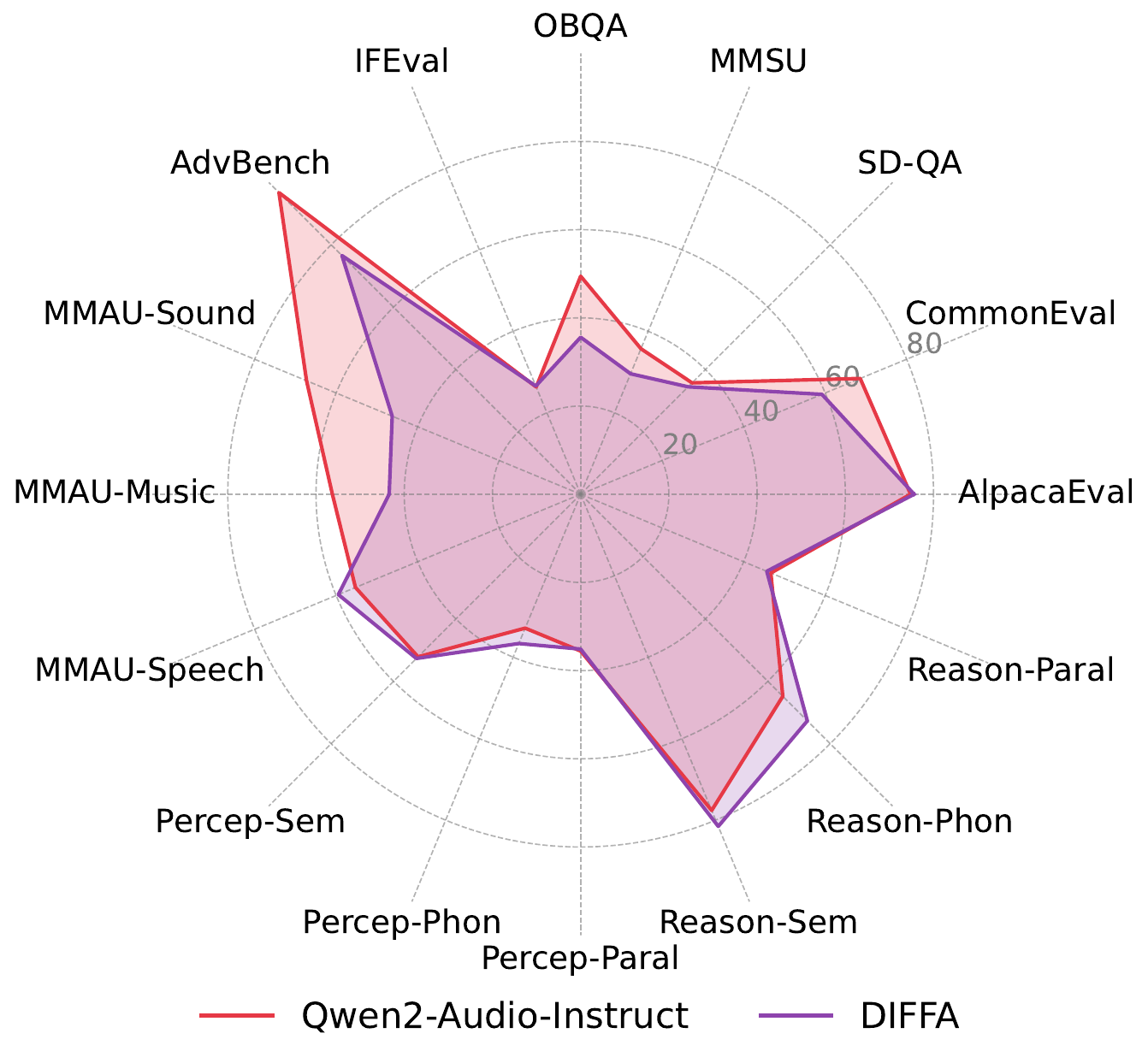}
\caption{DIFFA vs. Qwen2-Audio-Instruct. The abbreviations correspond to MMSU benchmark's capabilities: Perception-Semantics (Percep-Sem), Perception-Phonology (Percep-Phon), Perception-Paralinguistics (Percep-Paral), Reasoning-Semantics (Reason-Sem), Reasoning-Phonology (Reason-Phon), and Reasoning-Paralinguistics (Reason-Paral).}
\label{pic:radar}
\end{figure}

\section{Introduction}
Large language models (LLMs) have catalyzed a paradigm shift in artificial intelligence, pushing the frontiers of natural language understanding, computer vision, and multimodal reasoning~\cite{Gpt-4o}. In the domain of speech and audio processing, large audio-language models (LALMs) have similarly benefited from these advances in LLMs. By bridging continuous acoustic signals with discrete linguistic representations, LALMs enable end-to-end modeling of spoken interaction~\cite{Qwen-Audio,zhang2023speechgpt}. This capability not only advances fundamental research in speech understanding and generation, but also opens up practical opportunities for building more natural, robust, and versatile human-computer communication systems.

Existing LALMs typically follow two design paradigms. The first couples a speech encoder with an LLM, often through lightweight adapters that project continuous acoustic representations into the input space of the language model (e.g., Qwen2-Audio~\cite{Qwen2-Audio}, Audio-Flamingo~\cite{kong2024audio}). 
The second discretizes audio into speech tokens via speech tokenizers and subsequently trains directly on these tokens under the LLM training paradigm (e.g., SpeechGPT~\cite{zhang2023speechgpt}, Moshi~\cite{defossez2024moshi}).
Despite their strong results, both paradigms predominantly rely on autoregressive (AR) decoding, which suffers from well-known drawbacks such as exposure bias, slow generation, and limited flexibility for bidirectional or partially conditioned inference.

To address these limitations, diffusion-based language models\cite{Austin_Johnson_Ho_Tarlow_Berg_2021,shi2024simplified} have emerged as a promising alternative. By framing generation as an iterative denoising process, diffusion models support non-autoregressive decoding, parallel prediction, and improved controllability\cite{shi2024simplified}. Recent advances such as LLaDA~\cite{nie2025large} demonstrate that diffusion LLMs can rival autoregressive counterparts like LLaMA-3~\cite{dubey2024llama}, while exhibiting stronger robustness and training efficiency. Furthermore, LLaDA-V~\cite{you2025llada} extends this paradigm to vision–language tasks, confirming the competitiveness and generality of diffusion modeling in multimodal learning.

However, the audio modality remains notably underexplored in the context of diffusion-based language models. While such models have demonstrated promising results in text domains, their applicability to audio-language understanding has not been systematically investigated. The unique characteristics of audio—such as acoustic variability, complex temporal structures, and rich paralinguistic information—motivate an exploration into whether diffusion LLMs can be effectively extended to this domain with their flexible decoding mechanisms and bidirectional context modeling.

To bridge this gap, we explore the potential of adapting large diffusion-based language models for audio understanding. Specifically, we investigate whether such models can effectively process audio inputs and perform on par with, or surpass, strong autoregressive LALMs. Toward this goal, we introduce \textbf{DIFFA}, a \textbf{DIFF}usion-based large \textbf{A}udio-language framework.
DIFFA adopts a modular and efficient design: a pretrained speech encoder (Whisper~\cite{radford2023robust}), two lightweight adapters (semantic and acoustic), and a frozen diffusion-based LLM. To avoid catastrophic forgetting, we train only the adapters and keep both the language model and speech encoder frozen. The training procedure is divided into two stages: first, we align the semantic adapter under an ASR objective using LibriSpeech~\cite{panayotov2015librispeech}; then, we fine-tune both adapters on synthetic instruction data using the “What can you hear from the audio?” prompting scheme inspired by the DESTA-2~\cite{lu2025developing}. Figure~\ref{pic:radar} visually compares the performance of DIFFA and Qwen2-Audio-Instruct across multiple benchmarks. 
Notably, DIFFA attains competitive outcomes using merely 960 hours of ASR data and 127 hours of synthetic data, whereas Qwen2-Audio-Instruct depends on a far larger dataset of 510,000 hours.

Our contributions are summarized as follows:

\begin{itemize}
    \item We propose the first diffusion-based LALM, DIFFA, enabling large-scale audio-text understanding without relying on autoregressive modeling.
    \item We introduce a dual-adapter training framework and a two-stage training strategy that aligns speech representations to a frozen diffusion LLM, supporting audio understanding and instruction following without explicit supervised fine-tuning data.
    \item Despite using only 960 hours of ASR data, 127 hours of synthetic instruction data and 72 A800 GPU hours, DIFFA achieves competitive performance across multiple benchmarks, including MMSU, MMAU, and VoiceBench.
    \item We will release the training pipeline, inference code, and data generation scripts to promote research on diffusion-based LALMs with minimal compute and data requirements.
\end{itemize}

\section{Related Work}

\subsection{Large Audio-Language Models (LALMs)}

Large audio–language models (LALMs) have recently emerged as powerful tools for spoken language understanding in open-ended tasks. Existing approaches generally fall into two paradigms. The first couples a speech encoder with a pretrained LLM, often via lightweight adapters to bridge acoustic and textual representations. Representative examples include Qwen-Audio~\cite{Qwen-Audio} and Qwen2-Audio~\cite{Qwen2-Audio}, which integrate Whisper through a unified interface; SALMONN~\cite{tang2024salmonn}, which introduces a dual-encoder and window-level Q-Former to better capture long audio contexts; and Audio-Flamingo2~\cite{Audio-Flamingo-2}, which employs curriculum learning and dense alignment strategies. These systems achieve strong results on benchmarks such as MMAU and VoiceBench but remain tied to autoregressive decoding.

The second paradigm discretizes speech into tokens using quantizers or self-supervised encoders, and treats them as an additional input stream to the LLM (e.g., SpeechGPT~\cite{zhang2023speechgpt}, Moshi~\cite{defossez2024moshi}). While effective, this line of work also relies exclusively on autoregressive generation. To date, non-AR architectures such as diffusion models have not been explored for audio–language understanding.

\subsection{Diffusion-Based Language and Multimodal Models}

Diffusion-based models redefine generative modeling by reversing a corruption process that progressively masks or perturbs input data. Diffusion-LM~\cite{Austin_Johnson_Ho_Tarlow_Berg_2021,shi2024simplified,Sahoo2024SimpleAE} first applied this idea to discrete language modeling. LLaDA~\cite{nie2025large} subsequently scaled diffusion models to LLMs, showing strong performance on NLU and generation tasks. Notably, LLaDA employs a masked diffusion process with a principled likelihood objective and exhibits robustness in reasoning tasks. Its extension, LLaDA-V~\cite{you2025llada}, introduces a visual pathway for multimodal generation, achieving results competitive with autoregressive vision-language models.

These results motivate our exploration of diffusion-based modeling for audio, a modality that naturally benefits from bidirectional and non-sequential reasoning. Our work represents the first application of this framework to LALMs.

\subsection{Modality Alignment Without Explicit Supervised Fine-tuning}

A growing body of work seeks to eliminate reliance on human-annotated SFT datasets. BLSP~\cite{BLSP} and AudioChatLlama~\cite{audiochatllama} introduce similar \emph{behavior alignment}, where speech and text are treated as semantically equivalent inputs expected to elicit the same model output, but constrained in semantic alignment. DESTA~\cite{zhang2023speechgpt} and DESTA-2~\cite{lu2025developing} leverage synthetic instruction-tuning data, constructed by prompting LLMs with “What can you hear from the audio?” alongside speech metadata, to enable paralinguistic understanding. These strategies reduce the need for paired human supervision.
However, DESTA-2 typically follow a cascaded design, relying on Whisper for transcription before instruction response. In contrast, our model integrates both semantic and acoustic information via dual adapters and performs instruction-following in an end-to-end fashion using a frozen diffusion LLM.

\begin{figure*}[t]
\centering
\includegraphics[width=1.0\textwidth]{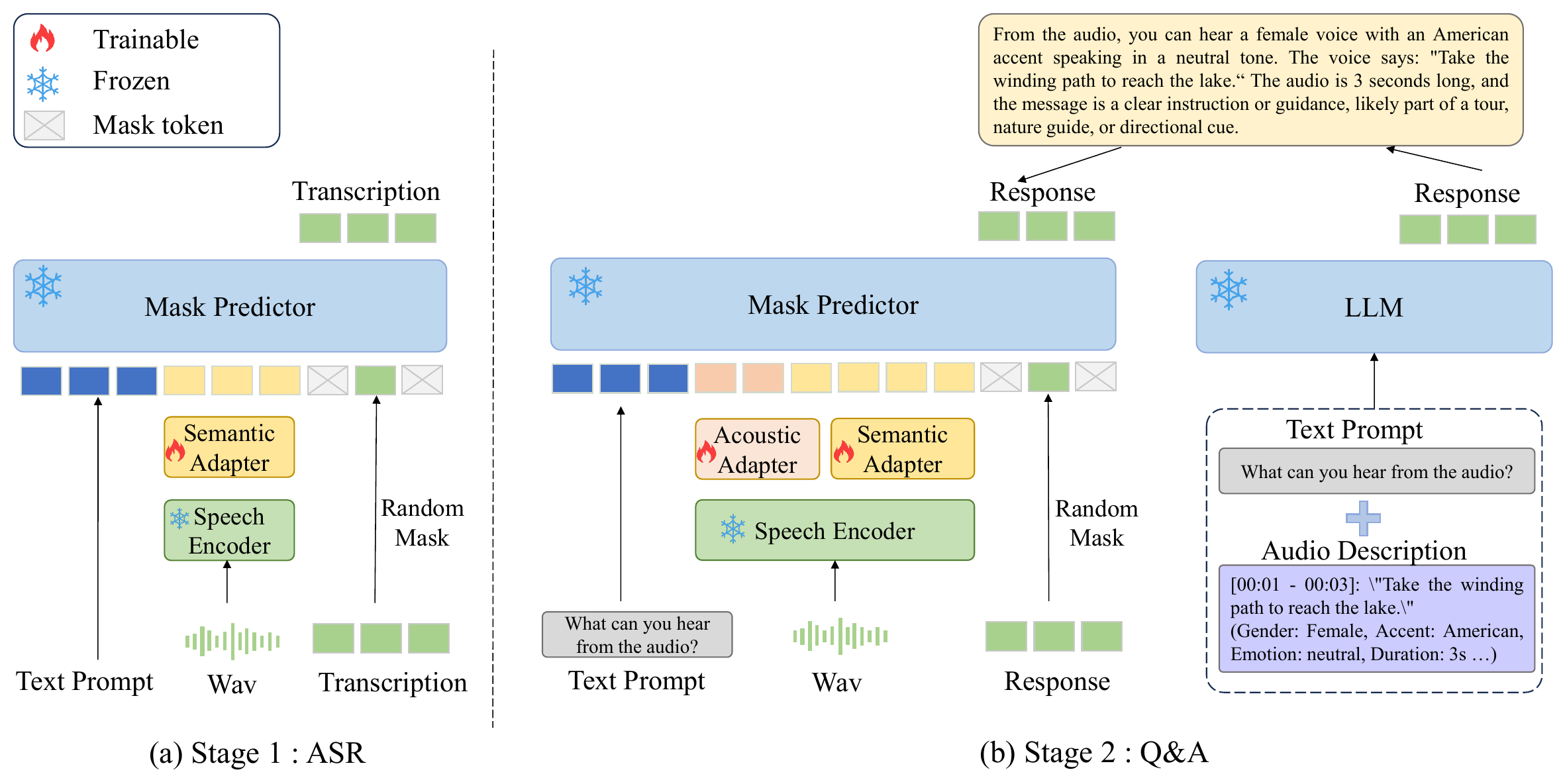}
\caption{Training process of our DIFFA framework. Stage 1 performs semantic alignment via an ASR objective, aligning the speech encoder with the diffusion language model. Stage 2 enables modality alignment by prompting the model to describe what it hears from the audio, following an audio caption instruction paradigm.}
\renewcommand{\arraystretch}{1.1}
\label{pic:train}
\end{figure*}

\section{Methods}
In this section, we present the overall framework of our proposed \textbf{DIFFA}, including its formulation, data construction, training strategy, and inference procedure.

\subsection{Preliminaries}
LLaDA~\cite{nie2025large} is a non-autoregressive language modeling paradigm that introduces a discrete random masking process and learns a \emph{mask predictor} to approximate its reverse. Unlike traditional autoregressive models, LLaDA allows for bidirectional dependency modeling and efficient likelihood-based training.

LLaDA defines a forward masking process to sample a corrupted sequence $x_t$, where each token is independently replaced with a special mask token $\textrm{M}$ with probability $t \in (0,1]$. The mask predictor $p_\theta(x_0 | x_t)$ is parameterized by a standard Transformer decoder and trained to reconstruct masked tokens:
\begin{equation}
\label{eq:pretrain-obj}
\mathcal{L}(\theta)  \triangleq   -  \mathbb{E}_{t, x_0,  x_t} \left[\frac{1}{t} \sum_{ i = 1 }^L \textbf{1}[x_t^i = \textrm{M}] \log p_{\theta}(x_0^i|x_t) \right] , 
\end{equation}
where $L$ denote the length of target sequence.
This objective yields a tractable upper bound of the negative log-likelihood~\cite{shi2024simplified,ou2025your}, enabling parallel token prediction.

Supervised fine-tuning (SFT) under LLaDA follows a similar approach. Given a prompt $p_0$ and response $r_0$, the response tokens are masked independently to obtain $r_t$. The loss is computed as:

\begin{equation}
\label{eq:sft-objective}
    - \mathbb{E}_{t, p_0, r_0, r_t} \left[\frac{1}{t} \sum_{i=1}^{L'} \textbf{1}[r_t^i = \textrm{M}] \log p_{\theta}(r_0^i|p_0, r_t) \right],
\end{equation}

where $L'$ is the response length.

During inference, LLaDA decodes iteratively from a fully masked sequence. At each denoising step, the model predicts masked tokens and re-applies masks to low-confidence positions, gradually refining predictions over $T$ steps.

\subsection{Data Construction}
Inspired by the DESTA series, we construct a dataset by prompting LLaDA-based or instruction-tuned language models (e.g., LLaMA3, Qwen3) with audio transcriptions: "\textit{[00:01 - 00:03]: "Take the winding path to reach the lake." 
(Gender: Female, Accent: American, Emotion: neutral, Duration: 3s …)
}" and acoustic attributes using the prompt: \textit{"What can you hear from the audio?"}. The generated response serves as the supervision signal, paired with the corresponding audio. This enables extbf{modality alignment without any explicit supervised fine-tuning data}. 

Furthermore, motivated by self-distillation techniques~\cite{yang-etal-2024-self}, we introduce a rewriting step to mitigate the domain shift arising from different model pre-training distributions. Specifically, we first employ Qwen3-8B to generate an initial set of captions. Subsequently, our LLaDA model rewrites these captions to align the textural style with its own internal data distribution. We denote this variant as LLaDA-rewrite-Qwen3.

\subsection{Model Architecture and Training Strategy}
Let $(a_0, p_0, r_0)$ denote the audio input, textual prompt, and target response, respectively. We employ a frozen Whisper-small encoder to extract frame-level acoustic features from $a_0$, and integrate them into the LLaDA-8B-Instruct backbone via two lightweight adapters.:

\textbf{Semantic Adapter.} A 2-layer convolutional network with a subsampling rate of 4, followed by a 2-layer linear projection. It compresses the encoder's 50 Hz output to 12.5 Hz.

\textbf{Acoustic Adapter.} A 2-layer Q-former~\cite{li2023blip} blocks with 64 trainable query vectors. It extracts speech-specific features from intermediate encoder states.

\textbf{Two-Stage Training.}
The whole training process is shown in Figure~\ref{pic:train}.
In stage 1, the semantic adapter is trained on 960 hours Librispeech using an ASR-style objective to align the speech encoder with the language model.
In stage 2, both adapters are fine-tuned on our 127-hour synthetic dataset under the audio captioning objective. The final audio representation is the concatenation of outputs from both adapters and prepended as prefix tokens to the LLM input. 
In both stages, audio and prompt tokens remain unmasked during training. We use <\texttt{endoftext}> as a padding and end-of-sequence token during training, which must also be predicted.
The LLaDA model and Whisper encoder remain frozen throughout. 
At each training step, the tokens of $r_0$ are independently replaced with a special mask token $\textrm{M}$ with probability $t \in (0,1]$. And then a forward masking process to sample a corrupted sequence $r_t$.
We optimize the model using a diffusion-style masked prediction objective:

\begin{align}
\label{eq:audio-train-loss}
L_{a} = & - \mathbb{E}_{t, a_0, p_0, r_0, r_t} \nonumber \\
& \left[\frac{1}{t} \sum_{i=1}^{L'} \mathbf{1}[r_t^i = \textrm{M}] 
\log p_{\theta}(r_0^i \mid a_0, p_0, r_t) \right],
\end{align}

where $r_t$ is the masked response and $L'$ is its length.

\subsection{Inference Procedure}
At inference time, we first pad the prompt and audio input, then initialize the response $r_T$ as a fully masked sequence of desired length. The model iteratively refines $r_t$ over $T$ denoising steps.

At step $t \rightarrow s$, the model predicts masked tokens:
\begin{equation}
\hat{r}_t = \arg\max p_\theta(r_0 | a_0, p_0, r_t),
\end{equation}
then re-masks $\lceil s/t \rceil$ proportion of tokens with the lowest confidence to form $r_s$.

\begin{figure}[t]
\centering
\includegraphics[width=0.5\textwidth]{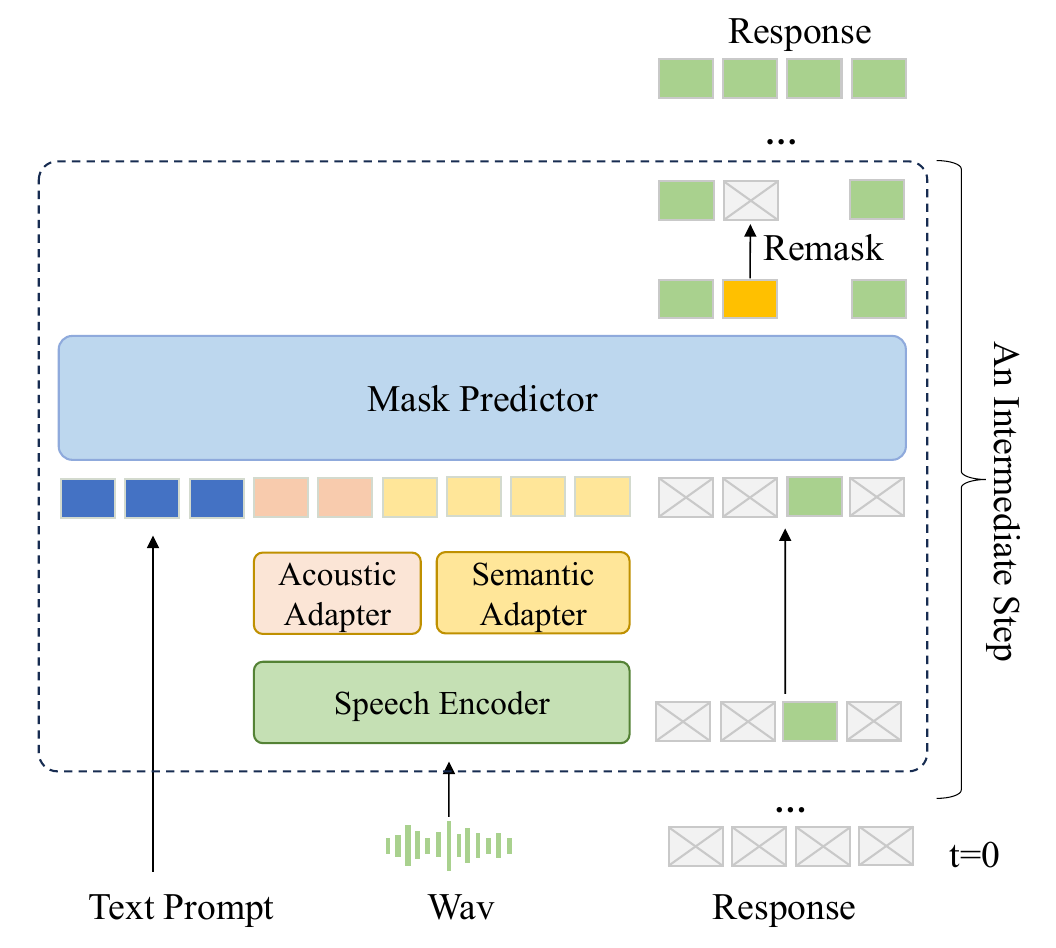}
\caption{Inference procedure of DIFFA.}
\label{pic:inference}
\end{figure}

We follow a semi-autoregressive strategy~\cite{nie2025large}, generating the sequence block-wise from left to right. Within each block, tokens are predicted in parallel and partially remasked.

This iterative inference scheme balances generation quality and efficiency, while maintaining the benefits of parallel decoding and bidirectional context modeling inherent in diffusion-based LLMs.

\section{Experimental Setup}
\subsection{Datasets}
In our experiments, we employ only open-source datasets Librispeech for ASR task in stage 1 and five dataset to construct dataset for Q\&A in stage 2: VCTK-Corpus~\cite{Yamagishi_Veaux_MacDonald_2019}, Accentdb~\cite{ahamad-etal-2020-accentdb}, IEMOCAP~\cite{Busso_Bulut_Lee_Kazemzadeh_Mower_Kim_Chang_Lee_Narayanan_2008}, dailytalk~\cite{dailytalk}, VoxCeleb~\cite{Nagrani_Chung_Zisserman_2017}.
The details of datasets are presented in Table~\ref{tab:dataset_stats}.

Compared to DESTA-2, our dataset follows a similar construction paradigm but excludes the \textit{PromptTTS} and \textit{Mixed Noise \& Reverb} subsets due to lack of access. Our dataset includes 10 annotated attributes—\textit{gender, age, accent, emotion, pitch, volume, speaking speed, duration, intent, and spoken text}—which is slightly fewer than the 12 attributes used in DESTA-2.

\begin{table}[t]
\centering
\begin{tabular}{lcc}
\toprule
\textbf{Dataset} & \textbf{Samples} & \textbf{Total Dur. (h)} \\ 
\midrule
VCTK-Corpus   & 20,000 & 19.91  \\
Accentdb      & 16,874 & 19.28  \\
IEMOCAP       & 20,000 & 24.82  \\
dailytalk     & 20,000 & 18.17  \\
VoxCeleb1     & 20,000 & 45.83  \\
\midrule 
\textbf{Total} & 96,874 & 127.01 \\ 
\bottomrule
\end{tabular}
\caption{Statistics of Datasets}
\label{tab:dataset_stats}
\end{table}

\begin{table*}[t]
\centering
\scriptsize
\renewcommand{\arraystretch}{1.05}
\setlength{\tabcolsep}{4pt}
\begin{tabular}{l|cccc|cccc|c}
\toprule
\multirow{2}{*}{\textbf{Models}}     & \multicolumn{4}{c|}{\textbf{Perception}}                  & \multicolumn{4}{c|}{\textbf{Reasoning}}                  & \textbf{Overall} \\ 
                            & \textbf{Semantics} & \textbf{Phonology} & \textbf{Paralinguistics} & \textbf{Avg}   & \textbf{Semantics} & \textbf{Phonology} & \textbf{Paralinguistics} & \textbf{Avg}   & \textbf{Avg}     \\ \midrule
Human                       & 87.10     & 94.32     & 92.88           & 91.24 & 82.16     & 87.60     & 89.12           & 86.77 & 89.72   \\ \midrule
Gemini-1.5-Pro~\cite{Gemini-1.5}              & 57.06     & \textbf{53.60}     & 31.23           & \textbf{46.10} & 79.47     & \textbf{83.46}     & 46.33           & 76.16 & \textbf{60.68}   \\
Qwen2.5-Omni~\cite{Qwen2.5-0mni}                & 55.12     & 37.33     & \textbf{39.35}           & 42.50 & \textbf{88.00}     & 81.37     & 48.36           & \textbf{79.83} & 60.57   \\
Kimi-Audio~\cite{Kimi-audio}                  & 57.64     & 42.30     & 35.74           & 43.52 & 81.77     & 76.65     & \textbf{55.22}           & 76.03 & 59.28   \\
MiniCPM~\cite{MiniCPM-O}                     & 56.56     & 34.05     & 36.48           & 40.54 & 80.71     & 74.72     & 46.71           & 73.57 & 56.53   \\
GPT-4o-Audio~\cite{Gpt-4o-card}                & \textbf{59.70}     & 41.56     & 21.44           & 39.67 & 80.83     & 78.74     & 26.25           & 71.96 & 56.38   \\
MERaLiON~\cite{MERaLiON}                    & 54.49     & 33.69     & 25.84           & 35.74 & 80.32     & 77.18     & 41.49           & 73.68 & 54.10   \\
Qwen2-Audio-Instruct~\cite{Qwen2-Audio}        & 52.14     & 32.87     & 35.56           & 39.02 & 77.62     & 64.81     & 46.67           & 68.90 & 53.27   \\
Gemini-2.0-Flash            & 47.17     & 41.30     & 30.62           & 40.83 & 70.69     & 70.69     & 36.16           & 47.83 & 51.03   \\
Megrez-3B-Omni~\cite{Megrez}              & 41.36     & 32.52     & 26.35           & 32.48 & 73.53     & 66.11     & 40.42           & 67.05 & 49.03   \\
DIVA\cite{DIVA}                        & 44.36     & 33.72     & 27.45           & 33.95 & 62.32     & 74.24     & 40.00           & 65.04 & 48.31   \\
Qwen-Audio-Chat~\cite{Qwen-Audio}             & 57.21     & 38.52     & 24.70           & 35.69 & 58.61     & 59.78     & 25.60           & 55.93 & 46.92   \\
Step-Audio~\cite{Step-Audio}                  & 31.56     & 29.39     & 24.01           & 28.72 & 49.10     & 50.09     & 45.27           & 47.27 & 37.42   \\
BLSP~\cite{BLSP}                        & 31.35     & 20.96     & 23.75           & 28.36 & 47.91     & 42.31     & 42.08           & 44.97 & 35.96   \\
GLM-4-Voice~\cite{GLM-4-Voice}                 & 27.80     & 24.52     & 27.34           & 26.18 & 46.10     & 48.16     & 44.35           & 46.76 & 35.51   \\
Random                      & 24.30     & 25.70     & 26.10           & 24.90 & 23.80     & 25.40     & 25.40           & 25.02 & 25.37   \\ \midrule
DIFFA         & 52.67     & 36.65     & 35.12           & 40.28 & 81.53     & 72.68     & 45.67           & 72.92 & 56.04   \\
\bottomrule
\end{tabular}
\caption{Performance breakdown on the MMSU benchmark across perception and reasoning dimensions.}
\label{tab:mmsu}
\end{table*}

\subsection{Model Configuration and Training Setup}

For the speech encoder, we adopt the Whisper-Small encoder, which contains 88.2 million parameters. As the language backbone, we use LLaDA-8B-Instruct, a large language model trained with a masked denoising objective inspired by diffusion-based frameworks. It is built upon a Transformer decoder architecture with 32 layers, 32 attention heads, a hidden size of 4096, and approximately 8.1 billion parameters. The architecture follows LLaMA~\cite{touvron2023llama,dubey2024llama}, with key modifications including RMSNorm~\cite{zhang2019root} for normalization, SwiGLU~\cite{shazeer2020glu} for non-linearity, and rotary position embeddings (RoPE)~\cite{su2024roformer} for positional encoding.

In our experiments, all parameters of LLaDA-8B-Instruct are frozen. We introduce lightweight trainable adapters to integrate audio features, following prior work on parameter-efficient multimodal learning. The semantic adapter contains 14.4 million parameters and the acoustic adapter 22.3 million. Training details are provided in Appendix~\ref{appendix:train_hyp}.

\subsection{Benchmarks}

\textbf{MMSU}~\cite{wang2025mmsu} is a large-scale benchmark aimed at evaluating the perception and reasoning capabilities of SpeechLLMs in authentic spoken language scenarios. It consists of 5,000 carefully curated audio-question-answer triplets across 47 diverse tasks, covering a wide spectrum of linguistic and paralinguistic phenomena—including phonetics, prosody, semantics, emotion, and speaker traits. Tasks are categorized into two main dimensions: perception (e.g., intonation detection, disfluency recognition) and reasoning (e.g., sarcasm detection, code-switch QA), thereby enabling a comprehensive assessment of models' ability to extract, interpret, and reason over fine-grained acoustic and linguistic cues.

\textbf{MMAU}~\cite{sakshi2025mmau} is a benchmark designed to evaluate advanced audio understanding through human-annotated multiple-choice questions paired with audio clips. It covers three core domains—speech, music, and environmental sounds—and targets 27 distinct skills that require complex reasoning and expert-level knowledge. Unlike benchmarks focused on low-level perception, MMAU emphasizes high-level cognitive abilities, challenging models to perform multi-step reasoning and knowledge retrieval grounded in audio inputs. In our experiments, we use the Test-mini split of MMAU for evaluation.

\textbf{VoiceBench~}\cite{chen2024voicebench} is a comprehensive benchmark designed to evaluate the capabilities of LLM-based voice assistants. It primarily consists of audio queries synthesized via text-to-speech (TTS) from existing text-based benchmarks, simulating realistic user interactions in spoken form.
It focuses on semantic understanding and supports structured evaluation along three dimensions: general knowledge, instruction following, and safety. These dimensions respectively assess the model's ability to answer questions, follow spoken constraints, and reject harmful prompts. 
By converting textual tasks into audio, VoiceBench offers a targeted and rigorous testbed for audio-language models.

\section{Experiments}

\subsection{Evaluation on MMSU}

To assess the advanced reasoning abilities of our diffusion-based model, we evaluate DIFFA on the \textbf{MMSU} benchmark, which assesses fine-grained spoken language understanding across perception (semantics, phonology, paralinguistics) and reasoning dimensions. As shown in Table~\ref{tab:mmsu}, our model achieves an average accuracy of 56.04\%, highlighting its ability to handle a wide range of linguistically grounded tasks.

Although the overall performance trails top proprietary models like Gemini-1.5-Pro (60.68\%), DIFFA outperforms many strong autoregressive baselines, such as Qwen2-Audio-Instruct (53.27\%) and Gemini-2.0-Flash (51.03\%). This competitive result, despite relying solely on synthetic supervision and lightweight adapters, supports the viability of diffusion-based approaches for nuanced speech understanding.

\begin{table*}[!t]
\centering
\small
\setlength{\tabcolsep}{5pt}
\renewcommand{\arraystretch}{1.05}
\begin{tabular}{lcccc}
\toprule
\textbf{Model} & \textbf{Sound} & \textbf{Music} & \textbf{Speech} & \textbf{Average} \\
\midrule
Gemini 2.5 Pro~\cite{Gemini-2.5}              & 75.08 & 68.26 & \textbf{71.47}  & \textbf{71.60} \\
Qwen2.5-Omni~\cite{Qwen2.5-0mni}                & \textbf{78.10} & 65.90 & 70.60  & 71.53 \\
Phi-4-multimodal~\cite{Phi-4-multimodal}            & 65.47 & 64.37 & 67.27  & 65.70 \\
Audio Flamingo 2 Reasoning~\cite{Audio-Flamingo-2}  & 75.98 & \textbf{74.25} & 43.54  & 64.59 \\
GPT-4o Audio~\cite{Gpt-4o-card}                & 64.56 & 56.29 & 66.67  & 62.51 \\
Audio Flamingo 2~\cite{Audio-Flamingo-2}            & 71.47 & 70.96 & 44.74  & 62.39 \\
Qwen2-Audio-Instruct~\cite{Qwen2-Audio}        & 67.27 & 56.29 & 55.26  & 59.61 \\
GPT-4o mini Audio~\cite{Gpt-4o-card}           & 50.75 & 39.22 & 69.07  & 53.01 \\
Gemini Pro v1.5~\cite{Gemini-1.5}             & 56.75 & 49.40 & 58.55  & 52.97 \\
M2UGen~\cite{M2UGen}                      & 43.24 & 37.13 & 33.33  & 37.90 \\
MusiLingo~\cite{MusiLingo}                   & 43.24 & 40.12 & 31.23  & 38.20 \\
SALMONN~\cite{SALMONN}                     & 41.14 & 37.13 & 26.43  & 34.90 \\
MuLLaMa~\cite{MuLLaMa}                     & 33.03 & 32.34 & 17.42  & 27.60 \\
GAMA-IT~\cite{GAMA}                     & 30.93 & 26.74 & 10.81  & 22.83 \\
GAMA~\cite{GAMA}                        & 31.83 & 17.71 & 12.91  & 20.82 \\
LTU~\cite{LTU}                         & 20.42 & 15.97 & 15.92  & 17.44 \\
Audio Flamingo Chat~\cite{Audio-Flamingo}         & 25.23 & 17.66 & 6.91   & 16.60 \\ \midrule
DIFFA        & 46.25 & 43.41 & 59.46  & 49.71 \\
\bottomrule
\end{tabular}
\caption{Evaluation results on the MMAU benchmark. Each model is assessed across three core audio domains: sound, music, and speech.}
\label{tab:mmaubench}
\end{table*}

\begin{table*}[!t]
\centering
\small
\setlength{\tabcolsep}{3pt}
\renewcommand{\arraystretch}{1.05}
\begin{tabular}{l|ccccccc|c}
\toprule
\textbf{Model}                            & \textbf{AlpacaEval} & \textbf{CommonEval} & \textbf{SD-QA} & \textbf{MMSU$^\ast$}  & \textbf{OBQA}  & \textbf{IFEval} & \textbf{AdvBench} & \textbf{Overall} \\
\midrule
GPT-4o-Audio~\cite{Gpt-4o-card}                & \textbf{4.78}       & \textbf{4.49}       & \textbf{75.50} & \textbf{80.25} & \textbf{89.23} & \textbf{76.02}  & 98.65    & \textbf{86.43}   \\
Kimi-Audio~\cite{Kimi-audio}                  & 4.46       & 3.97       & 63.12 & 62.17 & 83.52 & 61.10  & \textbf{100.00}   & 76.93   \\
Baichuan-Omni-1.5~\cite{Baichuan-Omni}           & 4.50       & 4.05       & 43.40 & 57.25 & 74.51 & 54.54  & 97.31    & 71.14   \\
GLM-4-Voice~\cite{GLM-4-Voice}                 & 3.97       & 3.42       & 36.98 & 39.75 & 53.41 & 25.92  & 88.08    & 55.99   \\
DiVA~\cite{DIVA}                        & 3.67       & 3.54       & 57.06 & 25.76 & 25.49 & 39.16  & 98.27    & 55.70   \\
Qwen2-Audio~\cite{Qwen2-Audio}                 & 3.74       & 3.43       & 35.72 & 35.72 & 49.45 & 26.33  & 96.73    & 55.34   \\
Step-Audio~\cite{Step-Audio}                  & 4.13       & 3.09       & 44.21 & 28.33 & 33.85 & 27.96  & 69.62    & 49.77   \\
LLaMA-Omni~\cite{LLaMA-Omni}                  & 3.70       & 3.46       & 39.69 & 25.93 & 27.47 & 14.87  & 11.35    & 37.50   \\
VITA~\cite{VITA}                        & 3.38       & 2.15       & 27.94 & 25.70 & 29.01 & 22.82  & 26.73    & 34.68   \\
Slam-Omni~\cite{Slam-Omni}                   & 1.90       & 1.79       & 4.16  & 26.06 & 25.27 & 13.38  & 94.23    & 33.84   \\
Mini-Omni2~\cite{Mini-Omni2}                  & 2.32       & 2.18       & 9.31  & 24.27 & 26.59 & 11.56  & 57.50    & 31.32   \\
Mini-Omni~\cite{Mini-Omni}                   & 1.95       & 2.02       & 13.92 & 24.69 & 26.59 & 13.58  & 37.12    & 27.90   \\
Moshi~\cite{Moshi}                       & 2.01       & 1.60       & 15.64 & 24.04 & 25.93 & 10.12  & 44.23    & 27.45   \\ \midrule
DIFFA         & 3.78       & 2.96       & 34.45 & 29.57 & 35.60 & 26.56  & 76.54    & 48.22    \\
\bottomrule
\end{tabular}
\caption{Evaluation results on VoiceBench. Metrics cover diverse QA and alignment tasks. Note that MMSU$^\ast$ in VoiceBench is derived from MMLU-Pro, which differs from the MMSU benchmark.}
\label{tab:voicebench}
\end{table*}

DIFFA performs particularly well on semantic reasoning tasks (81.53\%), benefiting from its strong language modeling backbone and ASR-aligned speech encoder. However, like most models in the benchmark, it exhibits lower accuracy in phonological and paralinguistic tasks—areas that demand precise acoustic perception beyond textual semantics. These trends mirror the broader challenges outlined in the MMSU benchmark, where human-level performance (89.72\%) remains a distant target.

This result provides the first empirical evidence that diffusion-based language models can serve as viable backbones for large-scale audio-language understanding, even without autoregressive decoding. Despite using only 127 hours of synthetic training data—orders of magnitude less than the tens of thousands of supervised fine-tuning (SFT) hours used by many baselines—DIFFA demonstrates strong generalization and reasoning capabilities.

Overall, DIFFA demonstrates robust generalization across complex linguistic dimensions, establishing a strong diffusion-based baseline. Future work should focus on enhancing the model’s sensitivity to prosodic and phonological cues to bridge the gap with human-level performance.

\begin{table*}[!t]
\setlength{\tabcolsep}{4pt}
\renewcommand{\arraystretch}{1.1}
\small
\centering
\begin{tabular}{ccc|cccc|ccc}
\toprule
\multirow{2}{*}{LLM Backbone} & \multirow{2}{*}{Data Source} & \multirow{2}{*}{Adapter} & \multicolumn{4}{c|}{MMAU}       & \multicolumn{3}{c}{MMSU}       \\ 
                          &                              &                          & Sound & Music & Speech & Avg   & Perception & Reasoning & Avg   \\ \midrule
LLaMA 3.1                 & LLaMA 3                      & Dual                     & 22.82 & 26.65 & 35.74  & 28.40 & 32.36      & 44.83     & 38.40 \\
LLaDA                     & LLaMA 3                      & Dual                     & \textbf{47.75} & \textbf{45.51} & \textbf{61.86}  & \textbf{51.71} & \textbf{37.27}      & \textbf{73.20}     & \textbf{54.72} \\ \midrule
LLaDA                     & Qwen3                        & Single                   & 44.44 & \textbf{44.31} & 54.35  & 47.70 & 36.98      & 69.83     & 52.88 \\ 
LLaDA                     & Qwen3                        & Dual                     & \textbf{46.25} & 43.41 & \textbf{59.46}  & \textbf{49.71} & \textbf{40.28}      & \textbf{72.92}     & \textbf{56.04} \\ \bottomrule
\end{tabular}
\caption{Ablation study on model architecture and adapter design. All models use 8B Instruct version.}
\label{tab:adapter-ablation}
\end{table*}

\begin{table}[!t]
\small
\setlength{\tabcolsep}{4pt}
\centering
\begin{tabular}{c|ccc|c}
\toprule
Data Source                 & MMAU  & MMSU  & Voicebench & Avg   \\
\midrule
LLaMA 3       & \textbf{51.71} & 54.72 & 37.17      & 47.86 \\
LLaDA         & 51.31 & 56.18 & 43.52      & 50.34 \\
Qwen3         & 49.71 & 56.04 & \textbf{48.22}      & \textbf{51.32} \\
rewrite-Qwen3 & 50.41 & \textbf{56.43} & 46.60      & 51.15 \\
\bottomrule
\end{tabular}
\caption{Ablation study on the impact of different instruction data sources. All models use 8B Instruct version.}
\label{tab:data-source-ablation}
\end{table}

\subsection{Evaluation on MMAU}

We further evaluate DIFFA on the MMAU benchmark, which tests 27 skills across three audio domains: sound, music, and speech. As shown in Table~\ref{tab:mmaubench}, DIFFA achieves an average accuracy of 49.71\%, outperforming several widely used autoregressive LALMs, such as SALMONN (34.90\%), GAMA-IT (22.83\%), and LTU (17.44\%). It also approaches the performance of commercial models like GPT-4o mini Audio (53.01\%) and Gemini Pro v1.5 (52.97\%).

A domain-level breakdown shows that DIFFA achieves the highest performance on speech-related tasks (59.46\%), likely benefiting from its speech-caption-focused training paradigm. In contrast, models such as Audio Flamingo 2 perform better on music and environmental sounds but underperform in speech understanding (e.g., 44.74\% for speech), underscoring the advantage of targeted, speech-centric training.

Overall, these results position DIFFA as a promising diffusion-based alternative to autoregressive LALMs. With limited resources and a parameter-efficient adapter tuning scheme, it achieves competitive results across complex audio reasoning benchmarks, indicating strong potential for scaling with larger or higher-quality datasets.

\subsection{Evaluation on VoiceBench}

VoiceBench evaluates semantic understanding from audio-as-question prompts, covering knowledge, instructions, and safety—making it a rigorous benchmark for spoken query comprehension.

Despite being trained on only 960 hours of ASR data and 127 hours of synthetic instructions, DIFFA achieves 34.45\% on SD-QA and 35.60\% on OBQA, demonstrating promising capability in factual spoken QA. This is particularly notable when compared to models like Qwen2-Audio (35.72\% SD-QA, 49.45\% OBQA), which are trained on hundreds of thousands of hours of proprietary data. The results suggest that diffusion-based models, even with limited training, can capture core semantic structures in speech.

On IFEval, DIFFA reaches 26.56\%, slightly surpassing Qwen2-Audio (26.33\%) and GLM-4-Voice (25.92\%), indicating a basic capacity for instruction comprehension from audio inputs. However, the performance gap with top models such as Kimi-Audio (61.10\%) underscores the challenge of aligning audio-conditioned instruction execution without large-scale supervised tuning.

In AdvBench, DIFFA attains 76.54\%, outperforming many strong baselines and approaching GLM-4-Voice (88.08). This highlights the potential of lightweight, diffusion-based models to learn safety-aligned behavior with minimal data.

In summary, DIFFA establishes a competitive baseline on VoiceBench despite using orders of magnitude less training data than competing systems. These findings validate the feasibility of diffusion-based LLMs for semantic audio understanding and offer a data-efficient alternative to current autoregressive paradigms.

\subsection{Ablation Study}

We perform a comprehensive ablation study to assess the impact of three key factors on model performance: (1) the choice of language model backbone, (2) adapter design, and (3) instruction data source. Besides, effect of inference hyperparameters are provided in Appendix~\ref{appendix:infer_hyp}.

\paragraph{Impact of Diffusion-Based Language Modeling.}  
As shown in Table~\ref{tab:adapter-ablation}, replacing the autoregressive LLaMA 3.1 backbone with the diffusion-based LLaDA architecture leads to a substantial improvement across all metrics. Specifically, with dual adapters, the LLaDA variant achieves 51.71 on MMAU and 54.72 on MMSU, significantly outperforming its LLaMA counterpart. This highlights the advantage of diffusion language model for audio understanding tasks.

\paragraph{Effect of Adapter Design.}  

We compare single and dual adapter configurations using the same LLaDA backbone. The single adapter setup uses only a semantic adapter aligned with the speech encoder's output. The dual adapter adds an acoustic adapter that extracts low-level acoustic cues from intermediate encoder states. Results show that dual adapters yield a +2.01 gain on MMAU and +3.16 on MMSU, confirming the advantage of combining semantic and acoustic information for richer audio understanding.

\paragraph{Instruction Data Source.}  
Table~\ref{tab:data-source-ablation} examines the role of different instruction data sources on final performance. All variants show comparable results, with Qwen3-generated instruction data leading to the best overall accuracy. Rewriting Qwen3 data with LLaDA brings only marginal improvements, suggesting that enhancing data quality at generation time may be more effective than post-hoc refinement. Interestingly, models trained on LLaDA-generated data outperform those based on LLaMA-3 instructions, indicating that the inductive bias of diffusion-based generation may yield more aligned supervision.

\section{Conclusion}
In this work, we introduce \textbf{DIFFA}, a diffusion-based large audio-language model that combines a frozen diffusion language backbone with lightweight dual adapters for audio understanding and instruction following. Using a two-stage training strategy and fully synthetic instruction data, DIFFA achieves competitive performance on diverse benchmarks—including MMSU, MMAU, and VoiceBench—despite relying on only 960 hours of ASR data and 127 hours of synthetic instruction data, in contrast to models like Qwen2-Audio-Instruct trained on over 500K hours.
As an initial exploration of diffusion-based modeling in the audio domain, our results suggest that such models offer a promising alternative to autoregressive LALMs. While our experiments are conducted on relatively small-scale open-source corpora, we plan to scale to broader and more diverse data in future work. We hope this study encourages further research into efficient, flexible, and controllable speech-driven AI systems.

\section*{Limitations}
A notable limitation is the constrained volume of training data. DIFFA is trained on a modest dataset comprising 960 hours of ASR data and 127 hours of synthetic instruction data, which may limit its capacity to generalize to a broader range of real-world audio phenomena—such as low-resource accents, noisy acoustic environments, or specialized domain speech. To address this, future iterations will focus on scaling up the training data, including expanding both ASR corpora and diverse synthetic instruction datasets, to strengthen the model’s robustness and coverage of complex audio scenarios.

\bibliography{main}

\appendix

\setcounter{figure}{0}  
\setcounter{table}{0}   
\renewcommand{\thefigure}{A.\arabic{figure}}
\renewcommand{\thetable}{A.\arabic{table}}

\section{Hyperparameter Settings}
\label{appendix:Hyperparameter}

\subsection{Training Hyperparameter}
\label{appendix:train_hyp}
In Stage 1, we train the semantic adapter using the LibriSpeech dataset for 10 epochs. We adopt a learning rate of 1e\textsuperscript{-4} with 1000 warm-up steps and a global batch size of 128. In Stage 2, both the semantic and acoustic adapters are jointly trained on our generated dataset for 10 epochs. We use a learning rate of 5e\textsuperscript{-5} with 2000 warm-up steps and a global batch size of 32. All experiments are optimized using the Adam optimizer and conducted on 4 NVIDIA A800 GPUs with 80GB memory each.

\subsection{Inference Hyperparameter}
\label{appendix:infer_hyp}

\begin{table}[!h]
\small
\begin{tabular}{cccc}
\toprule
Benchmark           & Answer length & Block length & Steps \\
           \midrule
MMSU       & 4             & 4            & 4     \\
MMAU       & 16            & 16           & 16    \\
\midrule
AlpacaEval & 128           & 32           & 128   \\
CommonEval & 128           & 32           & 128   \\
SD-QA      & 128           & 32           & 128   \\
MMSU       & 16            & 16           & 16    \\
OBQA       & 16            & 16           & 16    \\
IFEval     & 256           & 32           & 256   \\
AdvBench   & 128           & 32           & 128   \\
\bottomrule
\end{tabular}
\caption{Inference hyperparameters used for each benchmark}
\label{tab:hyperparams}
\end{table}

We detail the inference hyperparameters used in our experiments across different benchmarks in Table~\ref{tab:hyperparams}. Specifically, we configure the answer length (i.e., maximum number of tokens generated), block length (i.e., the number of tokens decoded per block from left to right), and the total number of denoising steps used in the diffusion-based decoding. For further details on the decoding process, please refer to LLaDA~\cite{nie2025large}. For decoding steps, we set it equal to answer length to achieve the best performance.

\begin{table}[!h]
\small
\begin{tabular}{cccc}
\toprule
Answer length & Block length & Steps & Alpacaeval \\ \midrule
128    & 16    & 128  & 3.72       \\
128    & 32    & 128  & \textbf{3.78}       \\
128    & 64    & 128  & 3.56       \\
128    & 128   & 128  & 3.07       \\ \midrule
256    & 16    & 256  & \textbf{3.78}       \\
256    & 32    & 256  & 3.76       \\
256    & 64    & 256  & 3.65       \\
256    & 128   & 256  & 3.28      \\
\bottomrule
\end{tabular}
\caption{Effect of answer length and block length on AlpacaEval performance}
\label{tab:ablation-block}
\end{table}

Effect of answer length and block length on AlpacaEval performance are reported in Table~\ref{tab:ablation-block} We observe that performance is sensitive to both the total answer length and the block size. Moderate block sizes (e.g., 32) consistently yield better results, likely due to a balance between context stability and denoising granularity.

\end{document}